\documentstyle[aps,eqsecnum,epsfig]{revtex}

\title{
\begin{flushright}\normalsize{hep-ph/0001067}\end{flushright}
Electromagnetic properties of a neutrino stream}
\author{Jos\'e F. Nieves} 
\address{
Laboratory of Theoretical Physics\\
Department of Physics, P.O. Box 23343\\
University of Puerto Rico, Rio Piedras, Puerto Rico}
\date{7 January 2000}

\begin{document}
\maketitle

\begin{abstract}

In a medium that contains a neutrino background 
in addition to the matter particles, the neutrinos
contribute to the photon self-energy as a result of
the effective electromagnetic vertex that they acquire
in the presence of matter.
We calculate the contribution to the photon self-energy
in a dense plasma, due to the presence of a gas of 
charged particles, or neutrinos,
that moves as a whole relative to the plasma.
General formulas for the transverse and longitudinal
components of the photon polarization tensor are obtained in terms
of the momentum distribution functions of the particles 
in the medium, and
explicit results are given for various limiting cases of
practical interest.  The formulas are used to study the 
electromagnetic properties of a plasma
that contains a beam of neutrinos.
The transverse and longitudinal photon dispersion relations
are studied in some detail.
Our results do not support the idea that
neutrino streaming instabilities can develop in such a system.
We also indicate how the phenomenon of optical activity of the neutrino
gas is modified due to the velocity of the neutrino background relative
to the plasma.
The general approach and results can be adapted to similar problems 
involving relativistic plasmas and high-temperature gauge theories
in other environments.

\end{abstract}

\section{Introduction and Conclusions}
\label{sec:intro}

From a modern point of view, the methods of finite temperature field 
theory (FTFT)\cite{ftft} provide a natural setting to treat the 
problems related to the propagation of photons and neutrinos
through a dense medium.  This view has been largely stimulated by the work of 
Weldon\cite{weldon:cov,weldon:fermions,weldon:imag}, who
emphasized the convenience of carrying out covariant, real-time calculations
in this kind of problem. The work of Weldon demonstrated that the real-time
formulation of FTFT is well suited to the study of systems involving 
gauge fields and/or chiral fermions at finite temperature, which can 
be extended in an efficient and transparent way to realistic situations
involving, for example, photons and/or neutrinos\cite{bterm} 
in a matter background.

The electromagnetic properties of neutrinos in a medium,
besides their intrinsic interest, 
are relevant in many physical applications\cite{raffeltbook}.
For example, the induced electromagnetic couplings of a neutrino 
propagating in a background of electrons and nucleons
is responsible for the plasmon decay process
$\gamma\rightarrow\nu\bar\nu$ in stars, and 
modify the MSW resonant condition in the presence of an 
external magnetic field\cite{dnp1,esposito-elmfors,pulsars,nuclmag}.
A neutrino gas also exhibits the phenomemon of optical
activity as a result of the chiral nature
of the neutrino interactions\cite{pisubpi,pisubpi2}.

The covariance in this type of calculation is implemented
by introducing the velocity four-vector $u^\mu$ of the medium,
in terms of which the thermal propagators are written in a manifestly
covariant form. Therefore, covariance is maintained, but quantities
such as the photon self-energy or the neutrino electromagnetic form factors
depend not just on the kinematical momentum variables of the problem,
but also on $u^\mu$.  In practice the vector $u^\mu$
is in the end set to $(1,\vec 0)$, which is equivalent to having carried
out the calculation from the start with respect to a frame of reference
in which the medium is at rest. This is usually
the relevant situation.
Therefore, while generally useful, the covariant nature of these methods
has not been of particular importance in the applications mentioned.

However, there is a class of problem in which setting $u^\mu = (1,\vec 0)$
is not possible. These are problems that involve one stream of particles
(which we can think of as a moving medium)
flowing through a background medium, which we can take to be stationary.
If we denote by $u^\mu$ the velocity four-vector of the stationary medium,
and by $u^{\prime\,\mu}$ the corresponding one for the moving background, then
we can set $u^\mu = (1,\vec 0)$, but
we cannot take both to be $(1,\vec 0)$ simultaneously. Thus, for example,
if we were to calculate the self-energy of the photon propagating
through such media, it will depend on the momentum and velocity
four-vector $u^\mu$ as usual, and in addition on $u^{\prime\,\mu}$. 
This additional dependence can have consequences
that are as important as the effects of the stationary background itself.

For example, it is well known that in a plasma in which a 
bunch of electrons move, as a whole, relative to a plasma
at rest, besides the usual dispersion relation of the longitudinal
photon mode, another branch appears whose dispersion relation
depends on the velocity of the beam. Under
some conditions, the sign of the imaginary part of this dispersion
relation is such that the corresponding amplitude grows exponentially,
which signifies an instability of the system against the excitation
of those modes.
This kind of system  
is familiar in plasma physics research, and examples 
of them are discussed in textbooks on the subject\cite{LLphyskin,ichimaru}.

Recently\cite{silvaetal}, it has been suggested that a similar kind
of streaming stability might be driven by a flow of neutrinos
through a matter background\cite{bento,footnote1}. 
Because the neutrino acquires an
effective electromagnetic coupling as it traverses a 
medium\cite{semikoz,sawyer,dnp1,nuclmag}, 
the propagation of a photon in a medium that contains a drifting neutrino
background may be affected in a way similar to the case mentioned above.
As argued in Ref.\ \cite{silvaetal}, such effects can appear under the conditions
of realistic situations such as those in a supernova explosion, 
gamma ray bursts, or the Early Universe.

Similarly, other neutrino processes that have been 
studied previously, such as those mentioned above, may be
modified in important ways if the neutrino gas is moving
as a whole relative to the matter background.

Motivated by all these considerations, in this work we calculate
the neutrino contribution to the photon self-energy in a medium
in which the neutrino gas moves as a whole relative to
a matter background which we take to be at rest.
The calculation is
based on the application of FTFT to this problem
in the manner that has been suggested above. The implicit assumption
is that, in the rest frame of the stream, the neutrino background
is characterized by a momentum distribution function that is
parametrized in the usual way. Although
our focus is the case in which 
the neutrino background constitutes the stream,
largely motivated by the potential applications that have been mentioned,  
the calculation and the formulas for the photon self-energy 
are presented in such a way that they can be adapted 
to other cases as well. Therefore, they
complement the existing calculations of the photon self-energy
in which all the particles form a common background with a unique
velocity four-vector.
The results for the photon self-energy can be equivalently
interpreted in terms of the dielectric constant 
of the system, and in that way we show that
the well known textbooks results for the stream stabilities
are reproduced when the appropriate limits are taken.
On the other hand, the results we obtain are valid
for general conditions (whether they are relativistic and/or degenerate
or the converse) of the gases that form the plasma at 
rest as well as the stream, hold
for general values of the velocity of the stream,
and are valid also for general 
values of the photon momentum and not necessarily for some particular limit.
Therefore, they are useful also in
the study of similar processes that may occur
in other contexts, such as high-temperature $QCD$,
heavy ion collisions or other similar environments in which
the methods of FTFT are applicable.

In Section \ref{sec:pif}, we give the general
one-loop formulas for the generic contribution of a 
moving fermion background to the photon self-energy.
The contribution from any given fermion can be written in terms of a few
independent functions, which are expressed as integrals over the
momentum distribution functions of the fermion.
Explicit formulas are given for various limiting cases of physical interest,
which also serve to show how some of the results derived 
in textbooks for simple cases are reproduced in the appropriate limit.

The case of the system that is composed of a matter 
background made of electrons and nucleons 
(and possibly their antiparticles), and a neutrino
gas that propagates as a whole relative to the matter background,
is considered in Section \ref{sec:pinu}.
We begin by reviewing the one-loop formulas
for the effective electromagnetic vertex of the neutrino in
a matter background, in the way that will be used in the 
calculation of the photon self-energy. The neutrino background 
contribution to the photon self-energy is then determined. It
depends on the momentum distribution functions of 
the matter particles and well as those for the neutrinos. 
As an application of the formulas obtained, 
the dispersion relation for the longitudinal
modes is considered, with attention to the possible effect of the neutrino
contribution to the instability of the system. In that context,
our results do not indicate the existence of unstable modes,
and therefore we do not find support for the idea that stream instabilities
due to the presence of the neutrino background can develop
in such systems.

In Section \ref{sec:pinu} we also consider the dispersion relations
for the transverse modes. The chiral nature of the neutrino interactions
gives rise to the phenomenon of optical activity, which had been
studied earlier\cite{pisubpi,pisubpi2}. Here we show how
the results of Refs.\ \cite{pisubpi,pisubpi2} are modified when the
neutrino gas is moving relative to the matter background. The main
effect is that the dispersion relations for the two
circularly polarized modes are not isotropic.  As a consequence,
the frequency splitting between them, which is the measure of the 
rotation of the plane of polarization, depends on the direction
of propagation of the mode relative to the velocity of the neutrino gas.
In particular, under the appropriate conditions,
the frequency difference 
is the opposite to what is found if the neutrino gas is not moving relative
to the matter background.

Section \ref{sec:conclusions} contains our outlook, where
other possible effects and applications are also mentioned.

%
%
\section{Photon self-energy in a fermion background}
\label{sec:pif}

\begin{figure}
\begin{center}
\epsfig{file=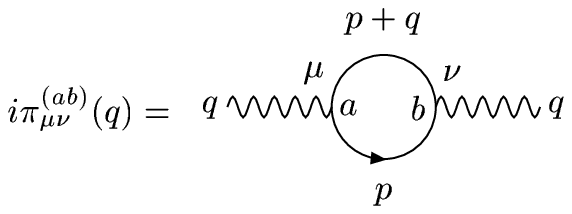}
\end{center}
\caption{Diagram for the contribution to the photon self-energy
matrix from a generic fermion. For a charged fermion, the electromagnetic
coupling is given by the tree-level terms in the Lagrangian while
for the neutrino it is the one-loop vertex function induced by the
matter background.
\label{fig:pimunu}}
\end{figure}

We will consider a medium that consists of a gas of nucleons, electrons,
neutrinos and their antiparticles.
Each fermion gas gives a contribution to the elements
of the $2\times 2$ photon self-energy matrix,
that are determined by calculating the diagram shown 
in Fig.\ \ref{fig:pimunu}. 
In particular, the contribution to the $\pi_{11\mu\nu}$ element from each
fermion $f$ in the loop is
\begin{equation}\label{pi11}
i\pi_{11\mu\nu}^{(f)} = (-1)(-i)^2 \mbox{Tr}\int\frac{d^4p}{(2\pi)^4}
j^{(em)}_{f\mu}(q) iS^{(f)}_{F11}(p + q)j^{(em)}_{f\nu}(-q) 
iS^{(f)}_{F11}(p) \,,
\end{equation}
where $j^{(em)}_{f\mu}(q)$ is the electromagnetic current of the fermion.
It is defined in such a way that the on-shell matrix element of the
electromagnetic current operator $J^{(em)}_\mu$ is given by
\begin{equation}\label{defj}
\langle f(p^\prime)|J^{(em)}(0)_\mu|f(p)\rangle = 
\bar u(p^\prime)j^{(em)}_{f\mu}(q)u(p) \,,
\end{equation}
where $q = p - p^\prime$ and $u(p)$ is a Dirac spinor.
For the electron it is simply $e\gamma_\mu$, for the nucleons
it must in principle include the magnetic moment term,
and for the neutrino we must use the effective electromagnetic
neutrino vertex in the medium.
The fermion propagator that appears in Eq.\ (\ref{pi11}) is given by
\begin{equation}\label{SF11}
S^{(f)}_{F11}(p) = ({p}\llap{/} + m_f)
\left[\frac{1}{p^2 - m_f^2 + i\epsilon}
+ 2\pi i \delta(p^2 - m_f^2)\eta_f(p\cdot u^{(f)})\right]
\end{equation}
where
\begin{equation}\label{etaf}
\eta_f(p\cdot u^{(f)}) = 
\frac{\theta(p\cdot u^{(f)})}{e^{\beta_f p\cdot u^{(f)} - \alpha_f} + 1}
+ \frac{\theta(-p\cdot u^{(f)})}{e^{-\beta_f p\cdot u^{(f)} + \alpha_f} + 1}\,,
\end{equation}
with $\beta_f$ being the inverse temperature and
$\alpha_f$ the fermion chemical potential.
The vector $u^{(f)\,\mu}$ is the velocity four-vector of the fermion gas
as a whole, so that $u^{(f)\,\mu} = (1,\vec 0)$ if 
the fermion background is at rest.
In Eq.\ (\ref{etaf}) we are allowing for the possibility 
that the different fermion gases of
the background may be at different temperatures and, most importantly
for our purposes later, that each gas has a velocity four-vector that
is not necessarily the same for all of them.
The implicit assumption here is that, in the rest frame of each
fermion background,
the corresponding particles have an isotropic thermal 
distribution characterized by a temperature and chemical 
potential $1/\beta_{f}$ and $\alpha_{f}$.

The dispersion relations of the propagating photon modes are obtained
by solving the equation
\begin{equation}\label{fieldeq}
(q^2 g_{\mu\nu} - q^\mu q^\nu - \pi^{(e\!f\!\!f)}_{\mu\nu})A^\nu = 0 \,,
\end{equation}
where 
\begin{equation}\label{repieffmunu}
\mbox{Re}\,\pi^{(e\!f\!\!f)}_{\mu\nu} = \sum_{f}\mbox{Re}\,\pi^{(f)}_{\mu\nu} \,,
\end{equation}
and we have denoted by $\pi^{(f)}_{\mu\nu}$ the 
background-dependent term of Eq.\ (\ref{pi11}).
In the rest of this paper we will focus only on the real part
of the dispersion relations, but similar considerations could
be used to calculate the imaginary part as well.

In order to calculate $\mbox{Re}\,\pi^{(e\!f\!\!f)}_{\mu\nu}$, and thus determine
the dispersion relations, we must know the composition of the
medium and the formulas for the electromagnetic current that
must be substituted in Eq.\ (\ref{pi11}). To proceed, we consider the various
cases separately.

\subsection{Matter background}

We consider first an isotropic medium composed of nucleon and electron
gases, with a common velocity four-vector $u^\mu$.
The most general form of the physical self-energy function in this case,
which we denote by $\pi^{(m)}_{\mu\nu}$ is\cite{weldon:cov,pisubpi}
\begin{equation}\label{pieffgen}
\pi^{(m)}_{\mu\nu} = \pi^{(m)}_T R_{\mu\nu}(q,u) + 
\pi^{(m)}_L Q_{\mu\nu}(q,u) + \pi^{(m)}_P P_{\mu\nu}(q,u) \,,
\end{equation}
where
\begin{eqnarray}\label{tensors}
R_{\mu\nu}(q,u) & = & g_{\mu\nu} - \frac{q^\mu q^\nu}{q^2} 
- Q_{\mu\nu}(q,u) \nonumber\\
Q_{\mu\nu}(q,u) & = & \frac{\tilde u_\mu\tilde u_\nu}{\tilde u^2}\nonumber\\
P_{\mu\nu}(q,u) & = & \frac{i}{{\cal Q}}
\epsilon_{\mu\nu\alpha\beta}q^\alpha u^\beta \,,
\end{eqnarray}
with
\begin{equation}\label{utilde}
\tilde u_\mu \equiv \left(g_{\mu\nu} - \frac{q_\mu q_\nu}{q^2}\right)
u^\nu \,.
\end{equation}
Although we have not indicated it explicitly,
in general, $\pi^{(m)}_{T,L,P}$ are functions of the scalar variables
\begin{eqnarray}\label{QOmega}
\Omega & = & q\cdot u \nonumber\\
Q & = & \sqrt{\Omega^2 - q^2} \,,
\end{eqnarray}
which have the interpretation of being the photon energy
and momentum in the rest frame of the medium.

A detailed calculation of the photon self-energy 
in such a medium was carried out in Ref.\ \cite{nuclmagpi}.  
For our present purposes,
it is useful to summarize those results as follows.
The nucleon magnetic moment term contribution
is not important for practical purposes. Therefore, we use here
$j^{(em)}_{f\mu} = e_f\gamma_\mu$, so that the neutron contribution
is being neglected.
Substituting in Eq.\ (\ref{pi11}) the formula for $S^{(f)}_{F11}$,
the contribution from each fermion in the loop
can be expressed in the form
\begin{eqnarray}\label{pimunuf}
\mbox{Re}\,\pi^{(f)}_{\mu\nu} & = & -4e^2\left[\frac{1}{2}\left(A_f(\Omega,{\cal Q}) - 
\frac{B_f(\Omega,{\cal Q})}{\tilde u^2}\right)R_{\mu\nu}(q,u)
+
\frac{B_f(\Omega,{\cal Q})}{\tilde u^2}Q_{\mu\nu}(q,u)\right]\,,
\end{eqnarray}
with the coefficients $A_f$ and $B_f$ defined as
\begin{eqnarray}\label{ABf}
A_f(\Omega,{\cal Q}) & = & \int\frac{d^3 p}{(2\pi)^3 2E_f}
\left(f_{f}(p\cdot u) + f_{\overline f}(p\cdot u)\right)
\left[
\frac{2m_f^2 - 2p\cdot q}{q^2 + 2p\cdot q} + 
(q\rightarrow -q)\right]\nonumber\\
B_f(\Omega,{\cal Q}) & = & \int\frac{d^3 p}{(2\pi)^3 2E_f}
\left(f_{f}(p\cdot u) + f_{\overline f}(p\cdot u)\right)
\left[
\frac{2(p\cdot u)^2 + 2(p\cdot u)(q\cdot u) - p\cdot q}
{q^2 + 2p\cdot q} + (q\rightarrow -q)\right]\,.
\end{eqnarray}
In these formulas,
\begin{equation}\label{pE}
p^\mu = (E,\vec p)\,, \quad E_f = 
\sqrt{{\vec p}^{\,2} + m_f^2} \,,
\end{equation}
and $f_{f,\overline f}$ denote the particle and antiparticle number density
distributions
\begin{equation}\label{ff}
f_{f,\overline f}({\cal E}) = \frac{1}{e^{\beta_f{\cal E} \mp \alpha_f} + 1}
\end{equation}
with the minus(plus) sign holding for the particles(antiparticles), 
respectively. 
Comparing Eqs.\ (\ref{pieffgen}) and (\ref{pimunuf}) we can identify
the contribution of any fermion to the
real part of the transverse and longitudinal components of the self-energy,
and therefore obtain
\begin{eqnarray}\label{repiTLmat}
\mbox{Re}\,\pi^{(m)}_T & = & -2e^2\sum_{f} \left(A_f(\Omega,{\cal Q})
+ \frac{q^2}{Q^2}B_f(\Omega,{\cal Q})\right) \,,\nonumber\\
\mbox{Re}\,\pi^{(m)}_L & = & 4e^2\sum_{f} \frac{q^2}{Q^2}B_f(\Omega,{\cal Q}) \,,
\end{eqnarray}
where the relation $\tilde u^2 = -{\cal Q}^2/q^2$ has been used.

\subsection{Matter background and a stream of charged particles}
\label{subsec:chargedstream}

We now consider a medium that contains, in addition to the
background as has been considered above, another gas of particles
with a velocity four-vector $u^\prime_\mu$.  We will refer to them
as the matter background and the stream, respectively,
and we assume that the latter consists of only one
specie of fermions $f^\prime$ with an electromagnetic
coupling $j^{(em)}_{f^\prime\mu} = e_{f^\prime}\gamma_\mu$.  
The fermion $f^\prime$ could be, for example, 
the electron or any other charged particle. We will denote
by $U^{\prime\,0}$ and $\vec U^{\prime}$ the components of $u^{\prime\,\mu}$
in the rest frame of the matter background so that, in that frame,
\begin{equation}\label{restframe}
u^\mu = (1,\vec 0) \,, \qquad 
u^{\prime\,\mu} = (U^{\prime\,0},\vec U^{\prime}) \,.
\end{equation}

The contribution from $f^\prime$ to the photon-self-energy
is given by a formula analogous to Eq.\ (\ref{pimunuf}),
\begin{equation}\label{pimunufprime}
 \pi^{(f^\prime)}_{\mu\nu} = \pi^{(f^\prime)}_T R_{\mu\nu}(q,u^\prime) +
 \pi^{(f^\prime)}_L Q_{\mu\nu}(q,u^\prime) \,,
\end{equation}
where
\begin{eqnarray}\label{piTLfprime}
\mbox{Re}\,\pi^{(f^\prime)}_T & = & -2e^2_{f^\prime}\left(
A_{f^\prime}(\Omega^\prime,{\cal Q}^\prime) + 
\frac{q^2}{{\cal Q}^{\prime\,2}}
B_{f^\prime}(\Omega^\prime,{\cal Q}^\prime)\right)\nonumber\\
\mbox{Re}\,\pi^{(f^\prime)}_L & = & 4e^2_{f^\prime}
\frac{q^2}{{\cal Q}^{\prime\,2}}B_{f^\prime}(\Omega^\prime,{\cal Q}^\prime)\,.
\end{eqnarray}
In Eq.\ (\ref{piTLfprime}) the functions $A_{f^\prime}$ and $B_{f^\prime}$ are given by
formulas analogous to Eq.\ (\ref{ABf}), but with the replacement
$u^\mu \rightarrow u^{\prime\mu}$, and we have used 
$\tilde u^{\prime\,2} = -{\cal Q}^{\prime\,2}/q^2$ where,
similarly to Eq.\ (\ref{utilde}),
\begin{equation}\label{uprimetilde}
\tilde u^\prime_\mu = \left(g_{\mu\nu} - \frac{q_\mu q_\nu}{q^2}\right)
u^{\prime\,\nu} \,.
\end{equation}
In analogy with Eq.\ (\ref{QOmega}), the variables $\Omega^\prime, {\cal Q}^\prime$
are defined by
\begin{eqnarray}\label{QOmegaprime}
\Omega^\prime & = & q\cdot u^\prime \nonumber\\
Q^\prime & = & \sqrt{\Omega^{\prime\,2} - q^2} \,,
\end{eqnarray}
and they are expressed in terms of $\Omega$ and ${\cal Q}$ by the relations
\begin{eqnarray}\label{QOmegarels}
\Omega^\prime & = & U^{\prime\,0}\Omega - \vec U^\prime\cdot\vec{\cal Q}
\nonumber\\[12pt]
{\cal Q}^\prime & = & \sqrt{\left(U^{\prime\,0}\Omega - 
\vec U^\prime\cdot\vec{\cal Q}\right)^2 - \Omega^2 + {\cal Q}^2} \,.
\end{eqnarray}
The total photon self-energy is given by
\begin{eqnarray}\label{piefftotal}
\pi^{(e\!f\!\!f)}_{\mu\nu} & = & \pi^{(m)}_T R_{\mu\nu}(q,u) + 
\pi^{(m)}_L Q_{\mu\nu}(q,u)
+ \pi^{(f^\prime)}_T R_{\mu\nu}(q,u^\prime) + 
\pi^{(f^\prime)}_L Q_{\mu\nu}(q,u^\prime) \,.
\end{eqnarray}

Eq.\ (\ref{piefftotal}) can be written in a convenient form by
the following procedure. 
From the definition of $R_{\mu\nu}$ given in Eq.\ (\ref{tensors})
we have
\begin{equation}\label{RRprimerel}
R_{\mu\nu}(q,u^\prime) = R_{\mu\nu}(q,u) + 
Q_{\mu\nu}(q,u) - 
Q_{\mu\nu}(q,u^\prime) \,.
\end{equation}
%
%
%
We now define the vectors
\begin{equation}\label{e12}
e^\mu_1 \equiv \frac{R^{\mu\nu}(q,u)u^\prime_\nu}
{\sqrt{N_1}} \,,\qquad
e^\mu_2 \equiv -iP^{\mu\nu}(q,u)e_{1\nu} \,.
\end{equation}
with
\begin{eqnarray}\label{N1}
N_1 & = & -u^\prime_\mu u^\prime_\nu R^{\mu\nu}(q,u) \nonumber\\
& = & \frac{(\tilde u\cdot u^\prime)^2}{\tilde u^2} - \tilde u^{\prime\,2}\,,
\end{eqnarray}
which can be expressed in the form
\begin{equation}\label{N1restframe}
N_1 = U^{\prime\,2}{\cal Q}^2 - \left(\vec U^\prime\cdot\vec{\cal Q}\right)^2 \,.
\end{equation}
It is easy to verify that $e^\mu_{1,2}$ are mutually orthogonal
and satisfy
\begin{equation}\label{e12prop}
e_{1,2}\cdot q = e_{1,2}\cdot\tilde u = 0 \,,\qquad e^2_{1,2} = -1 \,.
\end{equation}
Thus, together with $\tilde u^\mu$, they form a complete set
transverse to $q^\mu$, and therefore it is possible to express
$\tilde u^{\prime\mu}$ in terms of them. The desired relation,
which follows from substituting Eq.\ (\ref{tensors}) into Eq.\ (\ref{e12}), is
\begin{equation}\label{euprimerel}
\tilde u^\prime_\mu = \sqrt{N_1}e_{1\mu} + 
\left(\frac{\tilde u\cdot u^\prime}{\tilde u^2}\right)\tilde u_\mu \,,
\end{equation}
which substituting in the definitions given in Eq.\ (\ref{tensors})
yields the convenient formulas
\begin{eqnarray}\label{Qprime}
Q_{\mu\nu}(q,u^\prime) & = &
\frac{N_1}{\tilde u^{\prime\,2}} e_{1\mu} e_{1\nu} + 
\left(\frac{N_1}{\tilde u^{\prime\,2}} + 1\right)Q_{\mu\nu}(u,q) +
\sqrt{N_1}\frac{\tilde u\cdot u^\prime}{\tilde u^2 \tilde u^{\prime\,2}}
(e_{1\mu}\tilde u_\nu + \tilde u_\mu e_{1\nu}) \nonumber\\
P_{\mu\nu}(q,u^\prime) & = & 
\frac{{\cal Q} \tilde u\cdot u^\prime}{{\cal Q}^\prime\tilde u^2}
P_{\mu\nu}(q,u) + \frac{i{\cal Q}\sqrt{N_1}}{{\cal Q}^\prime\tilde u^2}
\left(\tilde u_\mu e_{2\nu} - \tilde u_\nu e_{2\mu}\right) \,.
\end{eqnarray}
On the other hand, as shown in Ref.\ \cite{pisubpi}, 
$R_{\mu\nu}$ can be decomposed in the form
\begin{equation}\label{Rdecomp}
R_{\mu\nu}(q,u) = -(e^\mu_1 e^\nu_1 + e^\mu_2 e^\nu_2) \,.
\end{equation}
In this way, using Eqs.\ (\ref{RRprimerel}) and (\ref{Qprime}) in Eq.\ (\ref{piefftotal})
together with the decomposition given in Eq.\ (\ref{Rdecomp}),
we finally arrive at
\begin{eqnarray}\label{piefftotalfinal}
\pi^{(e\!f\!\!f)}_{\mu\nu} & = & -e_{1\mu}e_{1\nu}
\left[\pi^{(m)}_T + \pi^{(f^\prime)}_T -
\frac{N_1}{\tilde u^{\prime\,2}}
\left(\pi^{(f^\prime)}_L - \pi^{(f^\prime)}_T\right)\right]
- e_{2\mu}e_{2\nu}\left(\pi^{(m)}_T + \pi^{(f^\prime)}_T\right)
\nonumber\\ \nonumber\\
& & \mbox{}
+ \frac{\tilde u_\mu\tilde u_\nu}{\tilde u^2}
\left[\pi^{(m)}_L + \pi^{(f^\prime)}_L 
+ \frac{N_1}{\tilde u^{\prime\,2}}
\left(\pi^{(f^\prime)}_L - \pi^{(f^\prime)}_T\right)\right]
+ \sqrt{N_1}\frac{(\tilde u\cdot u^\prime)}{\tilde u^2\tilde u^{\prime\,2}}
\left(\pi^{(f^\prime)}_L - \pi^{(f^\prime)}_T\right)
\left(e_{1\mu}\tilde u_\nu + \tilde u_\mu e_{1\nu}\right) \,.
\end{eqnarray}

Eq.\ (\ref{piefftotalfinal}), besides unfolding the main structure of the modes
in a particularly simple way, is a useful formula that allows us to obtain
the dispersion relation of the modes under a variety of conditions.
In the absence of the stream, the solutions consist of one longitudinal
mode with polarization vector $e^\mu_3 \propto \tilde u^\mu$,
and two degenerate transverse modes with polarization vectors
$e^\mu_{1,2}$ that satisfy $Q_{\mu\nu}e^\nu_{1,2} = 0$.
Their dispersion relations are determined by solving the equations
$q^2 = \mbox{Re}\,\pi^{(m)}_{L,T}$
for the longitudinal and transverse modes, respectively.
The presence of the stream breaks the degeneracy of the
transverse modes, and in general causes a mixing between them
with the longitudinal mode. In those cases in which it is
permissible to treat the mixing term (the last term in Eq.\ (\ref{piefftotalfinal}))
as a perturbation (e.g., the number density in the stream is sufficiently 
smaller than those in the matter background), 
the dispersion relations are obtained approximately by solving
\begin{eqnarray}\label{disprels}
q^2 & = &
\pi^{(m)}_T + \pi^{(f^\prime)}_T +
\frac{N_1 q^2}{{\cal Q}^{\prime\,2}}
\left(\pi^{(f^\prime)}_L - \pi^{(f^\prime)}_T\right)\nonumber\\
q^2 & = &
\pi^{(m)}_T + \pi^{(f^\prime)}_T\nonumber\\
q^2 & = &
\pi^{(m)}_L + \pi^{(f^\prime)}_L 
- \frac{N_1 q^2}{{\cal Q}^{\prime\,2}}
\left(\pi^{(f^\prime)}_L - \pi^{(f^\prime)}_T\right) \,.
\end{eqnarray}
with corresponding polarization vectors $e_{1,2}$ and $e_3\propto \tilde u$,
respectively. In Eq.\ (\ref{disprels}) we have used
the relation $\tilde u^2 = -{\cal Q}^2/q^2$ and the corresponding one for 
$\tilde u^{\prime\,2}$. 
If the mixing term is not sufficiently small so that the higher order terms
are important, then the full $2\times 2$ problem in the $e_1 - \tilde u$
plane must be considered which, although tedious, is straightforward.

In the equations Eq.\ (\ref{disprels}), it is understood that the variables 
$\Omega^\prime, {\cal Q}^\prime$ are to be expressed in terms of
$\Omega, {\cal Q}$  by means of the relations given in Eq.\ (\ref{QOmegarels}).
They thus become implicit equations for $\Omega, {\cal Q}$, whose
solutions determine the dispersion relations $\Omega({\cal Q})$
of the various modes.

\subsection{Discussion}
\label{subsec:discussion}

For illustrative purposes, we consider the specific case of a
stream of electrons and a 
matter background that consists of an electron gas
and a non-relativistic proton gas.
We borrow from Ref.\ \cite{nuclmag}[see Eqs. (A5) and (A9)] the following results
\begin{eqnarray}\label{ABfqsmall}
B_f(\Omega,{\cal Q}) & = & -\frac{1}{2}\int\frac{d^3{\cal P}}{(2\pi)^3}
\frac{\vec{\cal Q}\cdot\nabla_{{\cal P}}(f_f({\cal E}) + f_{\overline f}({\cal E}))}
{\Omega - \vec v_{\cal P}\cdot\vec{\cal Q}} \,,\nonumber\\
A_f(\Omega,{\cal Q}) & = & B_f(\Omega,{\cal Q}) + 
\frac{\Omega}{2}\int\frac{d^3{\cal P}}{(2\pi)^3}
\frac{\vec v_{{\cal P}}\cdot\nabla_{{\cal P}}(f_f({\cal E}) + f_{\overline f}({\cal E}))}
{\Omega - \vec v_{\cal P}\cdot\vec{\cal Q}} \,,
\end{eqnarray}
where ${\cal E} = \sqrt{\vec{\cal P}^2 + m_f^2}$,
$\nabla_{{\cal P}}$ is the gradient operator with respect to the 
momentum $\vec{\cal P}$ and
$\vec v_{{\cal P}} = \vec{\cal P}/{\cal E}$. 
As shown there, they are valid for values of $q$ such that
\begin{equation}\label{smallqdef}
q/\langle{\cal E}\rangle \ll 1 \,,
\end{equation}
where $\langle{\cal E}\rangle$ stands for a typical average value
of the energy of the fermions in the gas.  
For distribution functions that depend on $\vec{\cal P}$ only through
${\cal E}$, we can replace 
$\nabla_{{\cal P}} \rightarrow \vec v_{{\cal P}}\frac{\partial}{\partial{\cal E}}$
in Eq.\ (\ref{ABfqsmall}).
Several useful formulas
follow from Eq.\ (\ref{ABfqsmall}) in particular cases.
For example, if the fermions are relativistic,
\begin{eqnarray}\label{ABfqsmallur}
A_f(\Omega,{\cal Q}) & = & -3\omega^2_{0f}\nonumber\\
B_f(\Omega,{\cal Q}) & = &
-3\omega^2_{0f}\left(1 - \frac{\Omega}{2{\cal Q}}
\ln\left|\frac{\Omega + {\cal Q}}{\Omega - 
{\cal Q}}\right|\right) \,.
\end{eqnarray}
Eq.\ (\ref{ABfqsmallur}) holds for a degenerate or non-degenerate gas.
For the non-relativistic and non-degenerate case we use
\begin{eqnarray}\label{ABfqsmallnr}
A_f(\Omega,{\cal Q}) & = & -3\omega_{0f}^2 + 
\frac{{\cal Q}^2\omega_{0f}^2}{\Omega^2}\nonumber\\
B_f(\Omega,{\cal Q}) & = & \frac{{\cal Q}^2\omega_{0f}^2}{\Omega^2}\,,
\end{eqnarray}
which are valid if, in addition to Eq.\ (\ref{smallqdef}),
\begin{equation}\label{smallqdefnr}
\Omega \gg \bar v_f{\cal Q} \,,
\end{equation}
where $\bar v_f \equiv 1/\sqrt{\beta_f m_f}$ is a 
typical value of the velocity of the fermions in the gas.
The quantity $\omega_{0f}^2$, which is related to the plasma frequency 
in the gas, is given by
\begin{equation}\label{omega0f}
\omega_{0f}^2 = \int\frac{d^3{\cal P}}{(2\pi)^3 2{\cal E}}
(f_f({\cal E}) + f_{\bar f}({\cal E}))
\left[1 - \frac{{\cal P}^2}{3{\cal E}^2}\right] \,.
\end{equation}
In Eqs.\ (\ref{ABfqsmallur}) and (\ref{ABfqsmallnr}) we have used
its form in the relativistic (ER) limit and non-relativistic (NR) limits,
\begin{equation}\label{omega0fur}
\omega^2_{0f} = \left\{
\begin{array}{ll}\frac{1}{6\pi^2}\int_0^\infty d{\cal P}{\cal P}
(f_f({\cal P}) + f_{\bar f}({\cal P})) & \mbox{(ER)}\\[12pt]
\frac{n_f}{4m_f} & \mbox{(NR)} \,, \end{array}\right.
\end{equation}
where $n_f$ is the fermion number density in the frame in which
the background is at rest.
The corresponding formulas for the non-relativistic and degenerate case are
given in Ref.\ \cite{nuclmagpi}. 

Under most circumstances, the protons make a negligible contribution
to the dispersion relations.
The conditions under which those terms can be important
are given in Ref.\ \cite{nuclmagpi}.  Here we do not include
those special cases and therefore we will not consider the protons further. 
Lastly, we assume that the stream is not moving too fast as a whole, 
so that the term that is
proportional to $N_1$ in Eq.\ (\ref{disprels}) can be neglected, since
according to Eq.\ (\ref{N1restframe}) it is or the order of the velocity
squared of the stream. 

With these assumptions, the dispersion relations become
\begin{eqnarray}\label{disprelsmodel}
q^2 & = & -2e^2\left[
\left(
A_{e}(\Omega,{\cal Q}) + 
\frac{q^2}{{\cal Q}^2}B_{e}(\Omega,{\cal Q})\right) + 
\left(
A_{e^\prime}(\Omega^\prime,{\cal Q}^\prime) + 
\frac{q^2}{{\cal Q}^{\prime\,2}}B_{e^\prime}(\Omega^\prime,{\cal Q}^\prime)
\right)\right]\nonumber\\
q^2 & = & 4e^2\left[
\frac{q^2}{{\cal Q}^2}B_{e}(\Omega,{\cal Q}) + 
\frac{q^2}{{\cal Q}^{\prime\,2}}
B_{e^\prime}(\Omega^\prime,{\cal Q}^\prime)\right]\,,
\end{eqnarray}
for the transverse and longitudinal modes, respectively.
We now consider several cases separately.

\subsubsection{Non-relativistic matter and stream electrons}
\label{subsec:nrstream}

For the electrons in the matter background we use
Eq.\ (\ref{ABfqsmallnr}). 
Similarly, for the stream
\begin{eqnarray}\label{ABstreamnr}
A_{e^\prime}(\Omega^\prime,{\cal Q}^\prime) & = & -3\omega^2_{0e^\prime}
+ \frac{{\cal Q}^{\prime\,2}\omega_{0e^\prime}^2}{\Omega^{\prime\,2}}
\nonumber\\
B_{e^\prime}(\Omega^\prime,{\cal Q}^\prime) & = &
\frac{{\cal Q}^{\prime\,2}\omega_{0e^\prime}^2}{\Omega^{\prime\,2}} \,,
\end{eqnarray}
which, as we will see, are suitable for 
finding the long wavelength limit of the dispersion relations. 
Eq.\ (\ref{ABstreamnr}) can be used if the solution is such that
\begin{equation}\label{smallQprimecond}
\Omega^\prime \gg \bar v_{e^\prime}{\cal Q}^\prime \,.
\end{equation}
The conditions under which the solution thus found is valid
can be ascertained afterwards. 
The formula for $\omega_{0e^\prime}^2$ is the same expression given
in Eq.\ (\ref{omega0f}), but with replacement
$f_{f,\bar f}({\cal E})\rightarrow f_{e^\prime,\bar e^\prime}({\cal E})$,
where
\begin{equation}\label{feprime}
f_{e^\prime,\bar e^\prime}({\cal E}) = 
\frac{1}{e^{\beta_{e^\prime}{\cal E} \mp \alpha_{e^\prime}} + 1} \,.
\end{equation}
As we have indicated previously, the implicit assumption that is being
made here is that the electrons that compose the stream have,
in the rest frame of the stream,
an isotropic thermal distribution characterized by a 
temperature and chemical $1/\beta_{e^\prime}$ and $\alpha_{e^\prime}$.

Let us consider the dispersion relation for
the longitudinal mode.
From Eqs.\ (\ref{repiTLmat}) and (\ref{piTLfprime}) this yields
\begin{equation}\label{Ldisprelexample2}
1 = 4e^2\left(\frac{\omega_{0e}^2}{\Omega^2} +
\frac{\omega_{0e^\prime}^2}{\Omega^{\prime\,2}}\right) \,.
\end{equation}
From Eq.\ (\ref{QOmegaprime}) 
\begin{equation}\label{Omegaprimeex}
\Omega^\prime = \Omega - \vec{\cal Q}\cdot\vec U^\prime
\end{equation}
using $U^{\prime\,0}\simeq 1$, and therefore 
the dispersion relation is
\begin{equation}\label{Ldisprelexample2a}
(\Omega - \vec{\cal Q}\cdot \vec U^\prime)^2 (\Omega^2 - 4e^2\omega^2_{0e})
- 4e^2\omega^2_{0e^\prime}\Omega^2 = 0 \,.
\end{equation}

The salient feature of Eq.\ (\ref{Ldisprelexample2a}) is that,
besides the usual solution 
$\Omega^2_L({\cal Q}\rightarrow 0) \simeq 4e^2\omega^2_{0e}$, there is another
one with $\Omega_L \simeq \vec U^{\prime}\cdot\vec{\cal Q}$. The standard
way to find this second solution is to substitute
\begin{equation}\label{longansatz}
\Omega =  \vec U^{\prime}\cdot\vec{\cal Q} + \delta_L
\end{equation}
in Eq.\ (\ref{Ldisprelexample2a}) and determine $\delta_L$ approximately by taking
it to be a small quantity. In this fashion, we find
\begin{equation}\label{longdelta}
\delta_L = \pm\frac{|2e\omega_{0e^\prime}\vec U^\prime\cdot\vec{\cal Q}|}
{\sqrt{(\vec U^\prime\cdot\vec{\cal Q})^2 - 4e^2\omega^2_{0e}}} \,,
\end{equation}
which shows the well-known instability of this system.
For values of $\vec U^\prime\cdot\vec{\cal Q}$ such that
\begin{equation}\label{instcond}
0 < |\vec U^\prime\cdot\vec{\cal Q}| < 2|e|\omega_{0e} \,,
\end{equation}
the dispersion relation has a solution with a positive imaginary part,
which signals that
the system is unstable against oscillations with those values of
$\vec U^\prime\cdot\vec{\cal Q}$.  The condition that $\delta_L$
be small relative to $\vec U^\prime\cdot\vec{\cal Q}$ is satisfied
for sufficiently small values of $\omega_{0e^\prime}/\omega_{0e}$.
On the other hand
notice that, for this solution, $\Omega^\prime = \delta_L$, and 
${\cal Q}^\prime = \sqrt{\Omega^{\prime\,2} - \Omega^2 + {\cal Q}^2}
\simeq {\cal Q}$.  The conditions given in
Eqs.\ (\ref{smallqdef}) and (\ref{smallQprimecond}) are then equivalent to
$|\vec U^\prime\cdot\vec{\cal Q}| \gg \bar v_{e}{\cal Q}$
and $\delta \gg \bar v_{e^\prime}{\cal Q}$ which,
for sufficiently small values of the thermal velocities,
are satisfied as well.

Turning now the attention to the transverse dispersion relation,
we substitute the formulas for $A_{e,e^\prime}$ and $B_{e,e^\prime}$
given in Eqs.\ (\ref{ABfqsmallnr}) and (\ref{ABstreamnr}) into Eq.\ (\ref{disprelsmodel}).
This yields simply
\begin{equation}\label{tdisprelnrstream}
q^2 = 4e^2\omega^2_{0e} + 4e^2\omega^2_{0e^\prime} \,,
\end{equation}
which shows that in this case the presence of the stream perturbs somewhat
the transverse dispersion relation by shifting the value of the
plasma frequency, but it does not produce a significant effect otherwise.

Eqs.\ (\ref{Ldisprelexample2}) and (\ref{tdisprelnrstream}) 
reproduce the well-known results found
in textbooks\cite{disprelsimple},
which are derived by kinetic theory or similar semi-classical methods.
However, the results that we have obtained, and which are summarized in
Eqs.\ (\ref{piefftotalfinal}) and (\ref{disprels}), go farther.  Together
with the expressions for the self-energy functions in terms
of the coefficients $A_f$ and $B_f$ [Eqs.\ (\ref{repiTLmat}) and (\ref{piTLfprime})] 
they allow us to study systems
under a wider variety of conditions, including those
for which the semi-classical
approaches, and the simple formula given in Eq.\ (\ref{Ldisprelexample2})
in particular, are not be applicable. 

\subsubsection{Non-relativistic matter electrons and
relativistic stream electrons}
\label{subsec:urstream}

For the electrons in the stream we must use in this case
\begin{eqnarray}\label{ABstreamur}
A_{e^\prime}(\Omega^\prime,{\cal Q}^\prime) & = & -3\omega^2_{0e^\prime} 
\nonumber \\
B_{e^\prime}(\Omega^\prime,{\cal Q}^\prime) & = &
-3\omega^2_{0e^\prime}\left(1 - \frac{\Omega^\prime}{2{\cal Q}^\prime}
\ln\left|\frac{\Omega^\prime + {\cal Q}^\prime}{\Omega^\prime - 
{\cal Q}^\prime}\right|\right) \,,
\end{eqnarray}
while the matter electron formulas are the same as the previous ones.
The dispersion relations are then determined by 
\begin{eqnarray}\label{lurstream}
\Omega^2 - 4e^2\omega^2_{0e} & = & 4e^2\omega^2_{0e^\prime}f_L \\[12pt]
\label{turstream}
\Omega^2 - {\cal Q}^2 - 4e^2\omega^2_{0e} & = & 4e^2\omega^2_{0e^\prime}f_T 
\end{eqnarray}
for the longitudinal and transverse modes, respectively,
where we have defined 
\begin{eqnarray}\label{fLT}
f_L & = & \frac{3\Omega^2}{{\cal Q}^{\prime\,2}}\left[
\frac{\Omega^\prime}{2{\cal Q}^\prime}\ln\left|\frac{\Omega^\prime
+ {\cal Q}^\prime}{\Omega^\prime - {\cal Q}^\prime}\right| - 1\right] 
\nonumber\\[12pt]
f_T & = &
\frac{3}{2}\left\{1 + \frac{q^2}{{\cal Q}^{\prime\,2}}
\left[1 - \frac{\Omega^\prime}{2{\cal Q}^\prime}
\ln\left|\frac{\Omega^\prime + {\cal Q}^\prime}{\Omega^\prime -
{\cal Q}^\prime}\right|\right]\right\} \,.
\end{eqnarray}

Let us consider the longitudinal dispersion relation.
\begin{figure}
\begin{center}
\epsfig{file=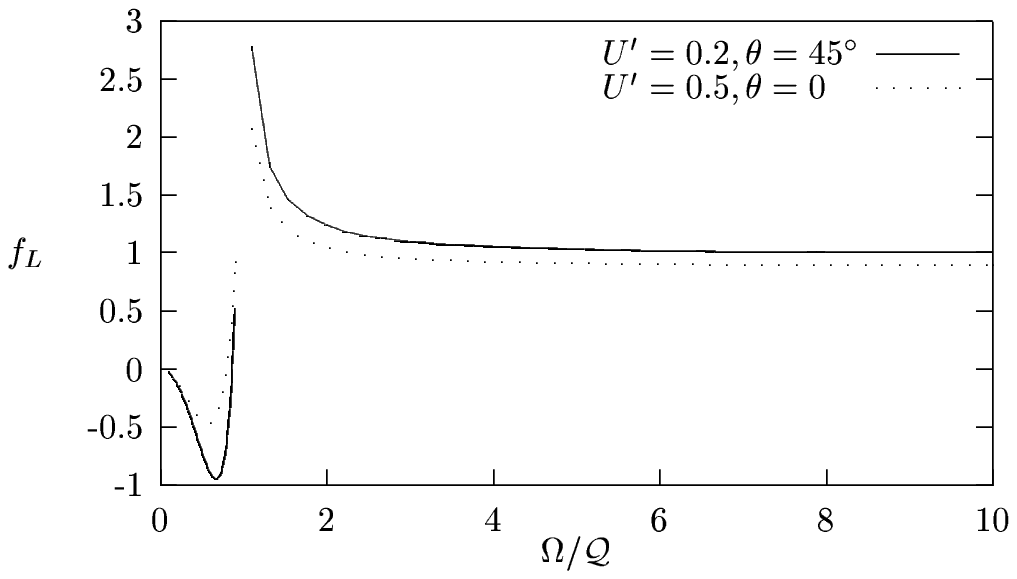}
\vspace*{0.5in}
{}
\epsfig{file=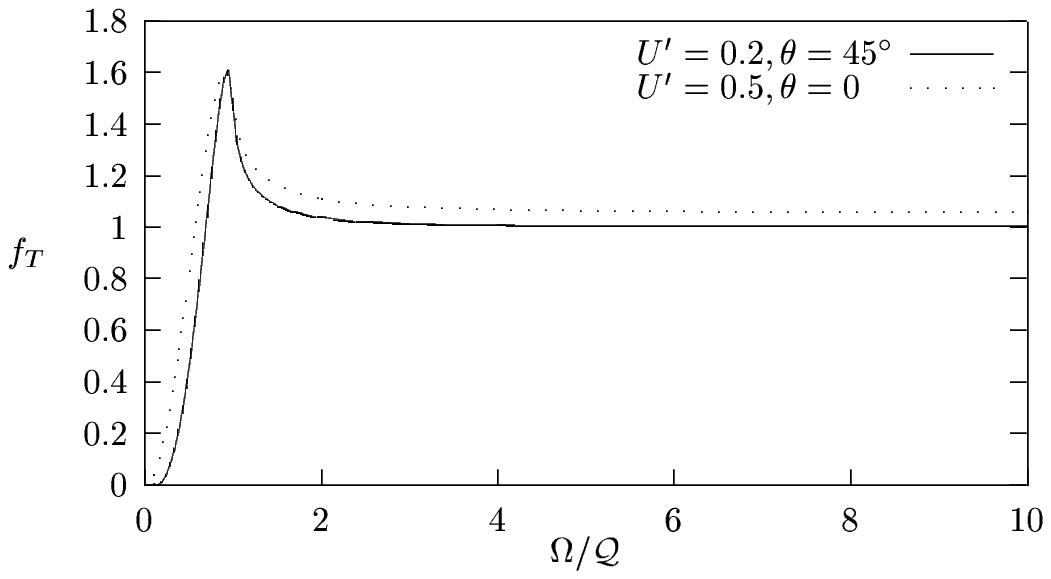}
\end{center}
\caption{The functions $f_{L,T}$ defined in the text, are plotted
as functions of $\Omega/{\cal Q}$, for some representative
values of the velocity of the stream $U^\prime$ and the angle between
$\vec{\cal Q}$ and $\vec U^\prime$. For $\Omega/{\cal Q} = 1$ the function
$f_L$ becomes infinite, while $f_T = 3/2$.
\label{fig:fLT}}
\end{figure}
With the help of Fig.\ \ref{fig:fLT} it is easy to see that,
besides the usual solution $\Omega_L \simeq 4e^2\omega_{0e}$,
any other solution to Eq.\ (\ref{lurstream}) must have $\Omega \approx {\cal Q}$
(which implies that $\Omega^\prime \approx {\cal Q}^\prime$ and the
function $f_L$ can be large)
so that the stream term can compete with the matter term in that equation.
Substituting 
\begin{equation}\label{luransatz}
\Omega = {\cal Q} + \delta_L
\end{equation}
in Eq.\ (\ref{lurstream}) we find
\begin{equation}\label{deltaLur}
\delta_L = \pm 2{\cal Q} e^{-2({\cal Q}^2 - 
4e^2\omega^2_{0e})/12e^2\omega^2_{0e^\prime}}
\end{equation}
for ${\cal Q}^2 > 4e^2\omega^2_{0e}$. In contrast to the
case considered in Section\ \ref{subsec:nrstream}, there is
no sign that a stream instability may develop in the present one.

For the transverse dispersion relations
the situation is different. The function $f_T$ is 
not larger than a number of order 1, as shown in Fig.\ \ref{fig:fLT},
so that the stream contribution in Eq.\ (\ref{turstream}) only 
introduces a perturbation in the usual solution.

In summary, when the stream consists of a relativistic electron gas,
there is no sign that any stream instabilities may develop.
This result will be a useful reference point when we consider
in Section\ \ref{sec:pinu}
the case in which the stream consists of neutrinos.

%
%
\section{Photon self-energy in a neutrino stream}
\label{sec:pinu}

In this section we consider a system composed of a matter background
as in Section\ \ref{subsec:chargedstream}, and a neutrino stream
with a velocity four-vector $u^{\prime\,\mu}$. 
Our immediate task is to determine
the neutrino stream contribution to the photon self-energy,
for which we must calculate a diagram similar to the one in
Fig.\ \ref{fig:pimunu}, but with a neutrino as the fermion in the loop.
Denoting the effective electromagnetic vertex of
the neutrino in the matter background by
$\Gamma^{(\nu)}_\mu(q)$, then 
\begin{equation}\label{pi11nu}
i\pi_{11\mu\nu}^{(\nu)} = (-1)(-i)^2 \mbox{Tr}\int\frac{d^4p}{(2\pi)^4}
\Gamma^{(\nu)}_{\mu}(q) iS^{(\nu)}_{F11}(p + q)\Gamma^{(\nu)}_{\nu}(-q) 
iS^{(\nu)}_{F11}(p) \,,
\end{equation}
where the neutrino propagator $S^{(\nu)}_{F11}$ is 
is given by Eq.\ (\ref{SF11}) (with $m_\nu$ = 0).  
The neutrino effective vertex can be expressed in the form
\begin{equation}\label{nuemvertex}
\Gamma^{(\nu)}_\mu(q) = T_{\mu\nu}(q)\gamma^\nu L\,,
\end{equation}
where $L = \frac{1}{2}(1 - \gamma_5)$ as usual, and
$T_{\mu\nu}$ can be decomposed as
\begin{equation}\label{Tmunu}
T_{\mu\nu} = T_T R_{\mu\nu}(q,u) + T_L Q_{\mu\nu}(q,u) + 
T_P P_{\mu\nu}(q,u) \,.
\end{equation}
A detailed calculation of the various terms in Eq.\ (\ref{Tmunu}) was carried
out in Ref.\ \cite{nuclmag}.  For our purposes here, we can summarize the main
results obtained there as follows.  

For practical purposes, the contribution to $T_{T,L}$ due to
the anomalous magnetic moment couplings of the nucleons
in the background is negligible, so that
\begin{eqnarray}\label{TTL}
T_T & = & 2\sqrt{2}|e|G_F a_p\left(
A_p(\Omega,{\cal Q}) - \frac{B_p(\Omega,{\cal Q})}{\tilde u^2}\right) + 
T^{(e)}_T \,,\nonumber\\
T_L & = & 4\sqrt{2}|e|G_F a_p\frac{B_p(\Omega,{\cal Q})}{\tilde u^2} + 
T^{(e)}_L \,,
\end{eqnarray}
while
\begin{equation}\label{TP}
T_P = T_P^{(e)} -4\sqrt{2}G_F b_p{\cal Q}(|e| + 2m_p\kappa_p)C_p(\Omega,{\cal Q}) -
8m_n\kappa_n\sqrt{2}G_F b_n{\cal Q} C_n(\Omega,{\cal Q}) \,.
\end{equation}
In these formulas 
\begin{eqnarray}\label{kappanp}
\kappa_p & = & 1.79\left(\frac{|e|}{2m_p}\right) \,,\nonumber\\
\kappa_p & = & -1.91\left(\frac{|e|}{2m_n}\right) \,,\nonumber\\
\end{eqnarray}
are the anomalous magnetic moment of the nucleons,
the coefficients $a_f$ and $b_f$ are the
neutral current couplings of the fermion $f$, while
$A_p$ and $B_p$ are defined in Eq.\ (\ref{ABf}) and
\begin{equation}\label{Cf}
C_f(\Omega,{\cal Q}) = \int\frac{d^3p}{(2\pi)^3 2E}
\left(\frac{\tilde u\cdot p}
{{\tilde u}^2}\right)
(f_{f} - f_{\bar f})\left[\frac{1}{q^2 + 2p\cdot q} + 
(q\rightarrow -q)\right]\,.
\end{equation}
The electron terms $T^{(e)}_X$ were calculated in Ref.\ \cite{dnp1}, 
and are given by
\begin{eqnarray}\label{Te}
T^{(e)}_T & = & 2\sqrt{2}eG_F\left(A_e(\Omega,{\cal Q}) - 
\frac{B_e(\Omega,{\cal Q})}{{\tilde u}^2}
\right)
\left\{\begin{array}{ll}
a_e + 1 & (\nu_e)\\
{}\\
a_e & (\nu_{\mu,\tau})
\end{array}
\right. \nonumber\\
T^{(e)}_L & = & 4\sqrt{2}eG_F\frac{B_e(\Omega,{\cal Q})}{{\tilde u}^2}
\left\{\begin{array}{ll}
a_e + 1 & (\nu_e)\\
{}\\
a_e & (\nu_{\mu,\tau})
\end{array}
\right. \nonumber\\
T^{(e)}_P & = & -4\sqrt{2}eG_F{\cal Q}C_e(\Omega,{\cal Q})
\left\{\begin{array}{ll}
b_e - 1 & (\nu_e)\\
{}\\
b_e & (\nu_{\mu,\tau})
\end{array}
\right. \,.
\end{eqnarray}

Substituting Eq.\ (\ref{Tmunu}) in Eq.\ (\ref{pi11nu}), we then obtain 
for the neutrino stream contribution to the photon self-energy
\begin{equation}\label{pimununu}
\mbox{Re}\,\pi^{(\nu)}_{\mu\nu} = -2T_{\mu\alpha}(q)T_{\nu\beta}(-q) 
J^{\alpha\beta}
\end{equation}
where
\begin{eqnarray}\label{Jdef}
J^{\alpha\beta} & \equiv & \int\frac{d^3p}{(2\pi)^3 2E}
\left\{\vphantom{\frac{q^2}{q^2}}\right.
\left(f_\nu(p\cdot u^\prime) + f_{\bar\nu}(p\cdot u^\prime)\right)
\left[\frac{2p^\alpha p^\beta - p^\alpha q^\beta - q^\alpha p^\beta + 
g^{\alpha\beta}p\cdot q}{q^2 - 2p\cdot q} + (q\rightarrow -q)\right]
\nonumber\\
& & \mbox{} + 
\left(f_\nu(p\cdot u^\prime) + f_{\bar\nu}(p\cdot u^\prime)\right)
i\epsilon^{\alpha\beta\lambda\rho}q_\lambda p_\rho
\left[\frac{1}{q^2 - 2p\cdot q} + \frac{1}{q^2 - 2p\cdot q}
\right]\left.\vphantom{\frac{q^2}{q^2}} \right\} \,.
\nonumber\\ 
\end{eqnarray}
The transversality and symmetry  properties of $J^{\alpha\beta}$ imply
that it is expressible in terms of the tensors
$R^{\alpha\beta}(q,u^\prime)$, $Q^{\alpha\beta}(q,u^\prime)$
and $P^{\alpha\beta}(q,u^\prime)$, with coefficients that
can be determined by projecting Eq.\ (\ref{Jdef}) along these tensors.
Thus we find
\begin{eqnarray}\label{defJ2}
J^{\alpha\beta} & = &
\frac{1}{2}\left(A_\nu(\Omega^\prime,{\cal Q}^\prime)
- \frac{B_\nu(\Omega^\prime,{\cal Q}^\prime)}{\tilde u^{\prime\,2}}\right)
R^{\alpha\beta}(q,u^\prime)\nonumber\\
& & \mbox{} +
\frac{B_\nu(\Omega^\prime,{\cal Q}^\prime)}{\tilde u^{\prime\,2}}
Q^{\alpha\beta}(q,u^\prime) +
{\cal Q}^\prime C_\nu(\Omega^\prime,{\cal Q}^\prime)
P^{\alpha\beta}(q,u^\prime) \,,
\end{eqnarray}
with the coefficients $A_\nu(\Omega^\prime,{\cal Q}^\prime)$,
$B_\nu(\Omega^\prime,{\cal Q}^\prime)$ and $C_\nu(\Omega^\prime,{\cal Q}^\prime)$
defined in Eqs.\ (\ref{ABf}) and (\ref{Cf}). In Eq.\ (\ref{defJ2})
the tensors $R_{\mu\nu}(q,u^\prime)$,
$Q_{\mu\nu}(q,u^\prime)$ and $P_{\mu\nu}(q,u^\prime)$ are
eliminated using Eqs.\ (\ref{RRprimerel}) and (\ref{Qprime}), and 
$R_{\mu\nu}(q,u)$ is decomposed as in Eq.\ (\ref{Rdecomp}).  
In this way,
\begin{eqnarray}\label{J}
J_{\alpha\beta} & = & -\left\{\frac{1}{2}\left(A_\nu - 
\frac{B_\nu}{\tilde u^{\prime\,2}}\right) -
\frac{N_1}{2\tilde u^{\prime\,2}}
\left[\frac{3B_\nu}{\tilde u^{\prime\,2}} - A_\nu\right]\right\}
e_{1\alpha}e_{1\beta} 
- \frac{1}{2}\left(A_\nu - 
\frac{B_\nu}{\tilde u^{\prime\,2}}\right)e_{2\alpha}e_{2\beta}
\nonumber\\[12pt]
& & \mbox{}
+ \left\{
\frac{B_\nu}{\tilde u^{\prime\,2}} + \frac{N_1}{2\tilde u^{\prime\,2}}
\left[\frac{3B_\nu}{\tilde u^{\prime\,2}} - A_\nu\right]
\right\}Q_{\alpha\beta}(q,u) +
C_\nu{\cal Q}\left(\frac{\tilde u\cdot u^\prime}{\tilde u^{\prime\,2}}\right)
P_{\alpha\beta}(q,u)\nonumber\\[12pt]
& & \mbox{} + 
\frac{\sqrt{N_1}\tilde u\cdot u^\prime}{2\tilde u^{\prime\,2}\tilde u^2}
\left[\frac{3B_\nu}{\tilde u^{\prime\,2}} - A_\nu\right]
\left(e_{1\alpha}\tilde u_\beta + \tilde u_\alpha e_{1\beta}\right)
+ \frac{iC_\nu\sqrt{N_1}{\cal Q}}{\tilde u^2}
\left(\tilde u_\alpha e_{2\beta} - e_{2\alpha}\tilde u_\beta\right) \,.
\end{eqnarray}
By substituting Eq.\ (\ref{J}) in Eq.\ (\ref{pimununu}) we finally obtain the formula
for the neutrino contribution which, when it is added to $\pi^{(m)}_{\mu\nu}$
to obtain the total photon self-energy, is the starting point
to determine the photon dispersion relations.  However, with all its 
generality, the formula is not particularly useful
and therefore we consider below some
specific situations of potential interest.

\subsection{Longitudinal dispersion relation}
\label{subsec:nulongitudinalmode}

As already seen in Section\ \ref{subsec:chargedstream}, 
in general the effects of the stream
break the degeneracy of the transverse modes and also mixes them
with the longitudinal one.  When the latter effect is not too large,
the longitudinal dispersion relation is obtained approximately by solving
the equation
\begin{equation}\label{longdisprelnu}
q^2 = \pi^{(m)}_L + \pi^{(\nu)}_l \,,
\end{equation}
where 
\begin{equation}\label{pinul}
\pi^{(\nu)}_l \equiv \frac{\tilde u^\mu\tilde u^\nu}{\tilde u^2} 
\pi^{(\nu)}_{\mu\nu} \,.
\end{equation}
Using the relation $\tilde u^\mu T_{\mu\alpha} = T_L\tilde u_\alpha$,
together with
\begin{eqnarray}\label{uproj}
\frac{\tilde u^\mu\tilde u^\nu}{\tilde u^2}R_{\mu\nu}(q,u^\prime) & = &
-\frac{N_1}{\tilde u^{\prime\,2}} \nonumber\\
\frac{\tilde u^\mu\tilde u^\nu}{\tilde u^2}Q_{\mu\nu}(q,u^\prime) & = &
\frac{N_1}{\tilde u^{\prime\,2}} + 1 \,,
\end{eqnarray}
we obtain from Eq.\ (\ref{pimununu})
\begin{equation}\label{pimununul}
\mbox{Re}\,\pi^{(\nu)}_l = 2T^2_L \frac{q^2}{{\cal Q}^{\prime\,2}}\left\{
B_\nu - N_1
\left[\frac{3q^2B_\nu}{{\cal Q}^{\prime\,2}} + 
\frac{1}{2}A_\nu\right]\right\} \,,
\end{equation}
where, to simplify the notation, we have omitted the arguments
$\Omega^\prime$ and ${\cal Q}^\prime$ in the coefficients $A_\nu$ and $B_\nu$.
The dispersion relation obtained by substituting Eqs.\ (\ref{repiTLmat}) and (\ref{pimununul})
in Eq.\ (\ref{longdisprelnu}) is the same as the corresponding one
for a stream of charged particles, with an effective charge
$e_\nu = \frac{1}{\sqrt{2}}T_L$.

\subsubsection{Long wavelength limit}

We consider the case analogous to the one discussed in 
Section\ \ref{subsec:discussion},
namely, a matter background made of a non-relativistic 
electron gas and a non-relativistic nucleon gas, 
and the neutrino stream.  As in the case mentioned, the photon
momentum is assumed to be such that Eq.\ (\ref{smallqdef}) holds. For
the electrons in the matter background we then use the formulas for
$A_e(\Omega,{\cal Q})$ and $B_e(\Omega,{\cal Q})$ given in Eq.\ (\ref{ABfqsmallnr}),
which are valid for $\Omega \gg \bar v_e {\cal Q}$ 
as indicated in Eq.\ (\ref{smallqdefnr}), 
while the analogous proton terms are negligible.  
From Eqs.\ (\ref{TTL}) and (\ref{Te}), this yields
\begin{equation}\label{TLsmallq}
T_L = - \frac{q^2}{\Omega^2}T_0 \,,
\end{equation}
where
\begin{equation}\label{T0}
T_0 \equiv 
4\sqrt{2}G_F e\omega_{0e}^2
\left\{\begin{array}{ll}
a_e + 1 & (\nu_e)\\
{}\\
a_e & (\nu_{\mu,\tau})
\end{array}
\right. \,.
\end{equation}
The neutrinos are, for all practical purposes,
ultrarelativistic.  For them, we use the formulas
analogous to those in Eq.\ (\ref{ABfqsmallur}), namely
\begin{eqnarray}\label{ABnuqsmall}
A_\nu(\Omega^\prime,{\cal Q}^\prime) & = & -3\omega^2_{0\nu}\nonumber\\
B_\nu(\Omega^\prime,{\cal Q}^\prime) & = &
-3\omega^2_{0\nu}\left(1 - \frac{\Omega^\prime}{2{\cal Q}^\prime}
\ln\left|\frac{\Omega^\prime + {\cal Q}^\prime}{\Omega^\prime - 
{\cal Q}^\prime}\right|\right) \,,
\end{eqnarray}
with
\begin{equation}\label{omega0nu}
\omega^2_{0\nu} = \frac{1}{6\pi^2}\int_0^\infty d{\cal P}{\cal P}
(f_\nu({\cal P}) + f_{\bar\nu}({\cal P})) \,,
\end{equation}
where
\begin{equation}\label{fnu}
f_{\nu,\bar\nu}({\cal P}) = \frac{1}{e^{\beta_\nu{\cal P} - \alpha_\nu} + 1}
\end{equation}
are the neutrino and antineutrino momentum distributions,
in the rest frame of the stream.  
The longitudinal dispersion relation obtained
from Eq.\ (\ref{longdisprelnu}) is
\begin{equation}\label{longdisprelnusmallq}
\Omega^2 = 4e^2\omega^2_{0e} + 2T^2_0 \frac{(q^2)^2}{\Omega^2{\cal Q}^{\prime\,2}}
\left\{B_\nu - N_1\left(\frac{3q^2}{{\cal Q}^{\prime\,2}}B_\nu
+ \frac{1}{2}A_\nu\right)
\right\} \,.
\end{equation}
In this equation, ${\cal Q}^\prime$ and $\Omega^\prime$ are
expressed in terms of ${\cal Q}$ and $\Omega$ by means of 
${\cal Q}^{\prime} = \sqrt{\Omega^{\prime\,2} - q^2}$ 
with $\Omega^\prime = U^{\prime\,0}\Omega - \vec U^\prime\cdot\vec{\cal Q}$,
as indicated by Eq.\ (\ref{QOmegaprime}).  
The solutions of Eq.\ (\ref{longdisprelnusmallq}) determine the
dispersion relation $\Omega_L({\cal Q})$ in the
long wavelength limit and are valid for $\Omega \gg \bar v_e {\cal Q}$.

\subsection{Neutrino driven stream instabilities}
\label{subsec:nuinstabilities}

Besides the usual solution $\Omega^2_L \simeq 4e^2\omega^2_{0e}$,
Eq.\ (\ref{longdisprelnusmallq}) can have an additional solution 
if the neutrino term is sufficiently large that it can compete with
the electron term. To determine whether this can happen, 
consider the specific situation in which the velocity of the neutrino stream
is not too large, so that the term in Eq.\ (\ref{longdisprelnusmallq})
proportional to $N_1$
can be neglected [see Eq.\ (\ref{N1restframe})].  Using the formula
for $B_\nu$ given in Eq.\ (\ref{ABnuqsmall}), the longitudinal
dispersion relation then becomes 
\begin{equation}\label{nuldisprel}
\Omega^2 - 4e^2\omega^2_{0e} = T^2_0\omega^2_{0\nu}
\left(1 - \frac{{\cal Q}^2}{\Omega^2}\right)^2 f_L \,,
\end{equation}
where $f_L$ is the same function defined in Eq.\ (\ref{fLT}).
For values of $\Omega \approx {\cal Q}$ the function $f_L$ becomes large,
but with the factor $(1 - {\cal Q}^2/\Omega^2)$ its contribution
in Eq.\ (\ref{nuldisprel}) is negligible in that region. 
On the other hand, for $\Omega\rightarrow 0$,
\begin{equation}
B_e(\Omega\rightarrow 0,{\cal Q}) = -\beta_e m_e\omega^2_{0e}
\end{equation}
so that
\begin{equation}
T_L = -T_0\beta_e m_e
\end{equation}
in this limit,
instead of Eq.\ (\ref{TLsmallq}). Therefore, the neutrino
contribution is not large in the limit $\Omega\rightarrow 0$ either.

We therefore conclude that the neutrino contribution
produces a small correction to the usual dispersion relation but does
not introduce any additional branch. 
In particular, there are no stream-induced instabilities in this system. 
This conclusion is in sharp contradiction with the result
obtained in Ref.\ \cite{silvaetal}, where it was found that, in the same system,
the dispersion relation indicates the appearance of
neutrino-driven stream instabilities.

To understand the origin of this discrepancy it is useful to consider
\begin{equation}\label{wrongdist}
f_{\nu,\bar\nu} = (2\pi)^2 n_{\nu,\bar\nu} 
\delta^{(3)}(\vec{\cal P} - {\cal E}\hat U^\prime)
\end{equation}
for the momentum distribution function of the neutrinos,
which is of the form used in Ref.\ \cite{silvaetal}.  Using
it in Eq.\ (\ref{ABfqsmall}) to calculate $B_\nu$ results in
\begin{equation}\label{Bnuqsmallwrong}
B_\nu = \left(\frac{n_\nu + n_{\bar\nu}}{2{\cal E}}\right)
\frac{{\cal Q}^2 - (\vec{\cal Q}\cdot\hat U^\prime)^2}
{(\Omega - \vec{\cal Q}\cdot\hat U^\prime)^2}\,,
\end{equation}
which, when  substituted in Eq.\ (\ref{longdisprelnusmallq}), 
yields the longitudinal dispersion relation
\begin{equation}\label{nuldisprelwrong}
\Omega^2(\Omega^2 - 4e^2\omega^2_{0e}) = 
\frac{T^2_0(n_\nu + n_{\bar\nu})}{{\cal E}}
\left(\frac{\Omega^2 - {\cal Q}^2}{{\cal Q}^{\prime}}\right)^2
\frac{{\cal Q}^2\sin^2\theta}
{(\Omega - {\cal Q}\cos\theta)^2}\,,
\end{equation}
where $\cos\theta = \vec{\cal Q}\cdot\hat U^\prime$ and
we have neglected the term proportional to $N_1$,
as before.  Eq.\ (\ref{nuldisprelwrong}) has a solution of the form 
\begin{equation}\label{OmegaLwrong}
\Omega_L = {\cal Q}\cos\theta + \delta^{(\nu)}_L  \,,
\end{equation}
with
\begin{equation}\label{deltaLnuwrong}
\delta^{(\nu)\,2}_L = \frac{T^2_0(n_\nu + n_{\bar\nu})}{{\cal E}}
\frac{{\cal Q}^2\sin^4\theta}
{\cos^2\theta({\cal Q}^2\cos^2\theta - 4e^2\omega^2_{0e})} \,,
\end{equation}
which exhibits an instability for ${\cal Q}^2\cos^2\theta < 4e^2\omega^2_{0e}$.

Thus, while we are able to reproduce qualitatively
the result of Ref.\ \cite{silvaetal}
in this way, we must note that it is based on an inconsistent application
of the long wavelength formulas given in Eq.\ (\ref{ABfqsmall}).
As explained in detail in Ref.\ \cite{nuclmagpi}, those formulas are obtained
by expanding the integrands in powers of $q/{\cal E}$
in the one-loop formulas given in Eq.\ (\ref{ABf}),
and retaining only those terms that are dominant in the limit
$q/{\cal E} \rightarrow 0$. This requires, among other conditions,
that the momentum distribution functions be such that
its derivatives do not introduce any singularities 
in the integrands.  The results derived in this way are equivalent
to those obtained by semiclasical methods based on kinetic theory
or similar approaches.  However, the form given in Eq.\ (\ref{wrongdist})
does not satisfy the required conditions and therefore neither
the long wavelength approximation of the one-loop formulas, 
nor the semiclassical formulas, are applicable in that case.

Leaving aside the question of whether or not a distribution function
such as that given in Eq.\ (\ref{wrongdist}) is realistic in any particular
physical context, in order to use it
the coefficients $A_\nu, B_\nu$ must be calculated
with the complete one-loop formulas given in Eq.\ (\ref{ABf}).
Following this procedure we obtain
\begin{equation}\label{Bnuother}
B_\nu =  2{\cal E}(n_\nu + n_{\bar\nu})\left[
\frac{{\cal Q}^2 - (\vec{\cal Q}\cdot\vec U^\prime)^2}
{4{\cal E}^2(\Omega - \vec{\cal Q}\cdot\vec U^\prime)^2 - (\Omega^2 - {\cal Q}^2)^2}
\right]
\end{equation}
instead of Eq.\ (\ref{Bnuqsmallwrong}),
so that the longitudinal dispersion relation becomes
\begin{equation}\label{nuldisprelright}
\Omega^2(\Omega^2 - 4e^2\omega^2_{0e}) = 
4T^2_0{\cal E}(n_\nu + n_{\bar\nu})
\left(\frac{\Omega^2 - {\cal Q}^2}{{\cal Q}^{\prime}}\right)^2
\frac{{\cal Q}^2\sin^2\theta}
{4{\cal E}^2(\Omega - {\cal Q}\cos\theta)^2 - (\Omega^2 - {\cal Q}^2)^2} \,.
\end{equation}
If we neglect here the $q^2$ term in the denominator,
we recover Eq.\ (\ref{nuldisprelwrong}). However, 
when that term is taken into account, the neutrino contribution
does not become large for 
$\Omega\approx\vec{\cal Q}\cdot\vec U^\prime$, and therefore
a solution of the form given in Eq.\ (\ref{OmegaLwrong}) does not exist.

\subsection{Transverse dispersion relation}

The dispersion relations for the transverse modes are given
approximately by solving the equation
\begin{equation}\label{tfieldeq}
\left[(q^2 - \pi^{(m)}_T)R_{\mu\nu}(q,u) - 
\pi^{(\nu)}_{t\mu\nu}\right]A^\nu = 0 \,,
\end{equation}
where
\begin{eqnarray}\label{pimununut}
\pi^{(\nu)}_{t\mu\nu} & \equiv & R_{\mu}{}^\alpha(q,u)R_{\nu}{}^\beta(q,u)
\pi^{(\nu)}_{\alpha\beta} \nonumber\\[12pt]
& = & -2\left[T_T(q)R_{\mu\lambda}(q,u) + T_P(q)P_{\mu\lambda}(q,u)\right]
\left[T_T(q)R_{\nu\rho}(q,u) - T_P(q)P_{\nu\rho}(q,u)\right]
J^{\lambda\rho}
\end{eqnarray}
is the transverse projection of the neutrino contribution
to the self-energy, and in the second line we have used 
Eq.\ (\ref{pimununu}).  For $J^{\alpha\beta}$ we will use the
formula given in Eq.\ (\ref{defJ2}) and consider the
case in which the terms with the factor $N_1$ can be neglected,
as we did in Section\ \ref{subsec:nuinstabilities}.
Therefore, remembering Eq.\ (\ref{Rdecomp}), we will substitute in Eq.\ (\ref{pimununut})
\begin{equation}\label{Japprox}
J^{\alpha\beta} = \frac{1}{2}\left(A_\nu - 
\frac{B_\nu}{\tilde u^{\prime\,2}}\right)R_{\alpha\beta}(q,u) 
+ C_\nu{\cal Q}\left(\frac{\tilde u\cdot u^\prime}{\tilde u^2}\right)
P_{\alpha\beta}(q,u)
\end{equation}

It is now useful to introduce the linear combinations\cite{footnote2}
\begin{equation}\label{Rpm}
R^{(\pm)}_{\alpha\beta} \equiv \frac{1}{2}\left(R_{\alpha\beta}(q,u)
\pm P_{\alpha\beta}(q,u)\right)\,,
\end{equation}
which satisfy 
\begin{equation}\label{Rpmorto}
R^{(s)\alpha\beta}R^{(s^\prime)}_{\beta\gamma} = 
\delta_{ss^\prime}\delta^\alpha_\gamma \,,
\end{equation}
and have the representation
\begin{equation}\label{Rpmrep}
R^{(\pm)}_{\alpha\beta} = - e^{(\pm)}_{\alpha} e^{(\pm)\dagger}_{\beta}
\end{equation}
where
\begin{equation}\label{chiralvectors}
e^{(\pm)}_{\alpha} = \frac{1}{\sqrt{2}}(e_{1\alpha} \pm ie_{2\alpha}) \,.
\end{equation}
Writing $R_{\alpha\beta}$ and $P_{\alpha\beta}$ in terms
of $R^{(\pm)}_{\alpha\beta}$ and substituting the resulting formulas
in Eq.\ (\ref{pimununut}), with the help of Eq.\ (\ref{Rpmorto}) we obtain
\begin{equation}\label{pimununut2}
\pi^{(\nu)}_{t\mu\nu} = \pi^{(+)} R^{(+)}_{\mu\nu} + 
\pi^{(-)} R^{(-)}_{\mu\nu}
\end{equation}
where
\begin{equation}\label{pipm}
\pi^{(\pm)} = -2\left(T_T \pm T_P\right)^2
\left[\frac{1}{2}\left(A_\nu - \frac{B_\nu}{\tilde u^{\prime\,2}}
\right) \pm C_\nu{\cal Q}\left(\frac{\tilde u\cdot u^\prime}
{\tilde u^2}\right)\right] \,.
\end{equation}
From Eq.\ (\ref{tfieldeq}), the dispersion relations for the transverse modes are then
\begin{equation}\label{tdisprels}
q^2 = \pi^{(m)}_T + \pi^{(\pm)}
\end{equation}
with the corresponding polarization vectors $e^{(\pm)}_\alpha$.

\subsection{Optical Activity of the neutrino gas}
\label{subsec:opticalactivity}

Eq.\ (\ref{pipm}) exhibits the phenomenon of optical activity of the neutrino
gas, in which the two circularly polarized photon modes travel
with different dispersion relations as a result of the chiral
interactions of the neutrino\cite{pisubpi}.  
Notice that for this occur,
$T_P$ and/or $C_\nu$ must be non-zero. This requires that 
the chemical potentials in the matter background be such that,
for some particle species, $\alpha_f \neq 0$, or that
$\alpha_\nu \neq 0$. In the latter case however, there is an additional
contribution to the photon self-energy that arises from the set
of diagrams that were calculated in detail in Ref.\ \cite{pisubpi2}.
Those diagrams are not included in Fig.\ \ref{fig:pimunu} and their result
is an additional contribution to the photon self-energy that
must be taken into account in Eq.\ (\ref{tdisprels}). 
The result of the calculation of Ref.\ \cite{pisubpi2} is taken into account
by including in the right-hand side of Eq.\ (\ref{pimununu}) the term
\begin{equation}\label{pisubpimunu}
\Pi^{(\nu)}_P P_{\mu\nu}(q,u^\prime)
\end{equation}
where, in the notation of the present paper,
\begin{equation}\label{pisubPnu}
\Pi^{(\nu)}_P = \frac{\sqrt{2}G_F \alpha}{3\pi}\frac{q^2}{m^2_e}
(n_\nu - n_{\bar\nu}){\cal Q}^\prime \,,
\end{equation}
with
\begin{equation}\label{nudensities}
n_{\nu,\bar\nu} = \int\frac{d^3{\cal P}}{(2\pi)^3} f_{\nu,\bar\nu}({\cal P}) \,.
\end{equation}
The result quoted in Eq.\ (\ref{pisubPnu}) is valid for values of $q < m_e$.
When this term is included in Eq.\ (\ref{pimununut}), the net effect
is that the right-hand side of Eq.\ (\ref{tdisprels}) has the additional the term
\begin{equation}\label{pisubpinu}
\pi_P = \pm \Pi^{(\nu)}_P\left(\frac{{\cal Q}}{{\cal Q}^\prime}\right)
\frac{\tilde u\cdot u^\prime}{\tilde u^2} \,.
\end{equation}
If the background contains an equal number of neutrinos and
antineutrinos, then $\Pi^{(\nu)}_P$ as well as $C_\nu$ are zero.
In such a case, the optical activity of the neutrino gas is due
to a non-zero value of $T_P$ in Eq.\ (\ref{pipm}), which in turn
depends on the difference between the particle and antiparticle
number densities in the matter background.
 
\subsubsection{Long wavelength limit}

$\pi^{(\pm)}$ can be evaluated explicitly by considering 
specific situations. As an example 
we consider once more a matter background that
consists of non-relativistic electron
proton gases, with the photon momentum satisfying Eq.\ (\ref{smallqdef})
and $\Omega \gg \bar v_f{\cal Q}$.
The proton contribution to $T_T$ is negligible, and using
Eq.\ (\ref{ABfqsmall}) for $A_e$ and $B_e$,
\begin{equation}\label{TTsmallq}
T_T = -T_0 \,,
\end{equation}
with $T_0$ given in Eq.\ (\ref{T0}).
On the other hand, $T_P$ is given by Eq.\ (\ref{TP}) where, in the momentum
regime that we are considering,
\begin{equation}\label{Cfqsmall}
C_f(\Omega,{\cal Q}\rightarrow 0) = -\frac{1}{2}\int
\frac{d^3{\cal P}}{(2\pi)^3 2{\cal E}}\left(\frac{f_f - f_{\bar f}}{{\cal E}}\right)
\left[1 - \frac{2{\cal P}^2}{3{\cal E}^2}\right]\,,
\end{equation}
as shown in Ref.\ \cite{nuclmag}. For the electron and proton gases
we use the non-relativistic limit of this,
$C_f = - \omega^2_{0f}/2m_f$,
which implies that the proton term is negligible and therefore
\begin{equation}\label{TPqsmall}
T_P = \left(\frac{{\cal Q}}{2m_e}\right)T^\prime_0 \,,
\end{equation}
where
\begin{equation}\label{Tprime0}
T^\prime_0 = 4\sqrt{2}G_F e\omega^2_{0e}
\left\{\begin{array}{ll}
b_e - 1 & (\nu_e)\\
{}\\
b_e & (\nu_{\mu,\tau}) 
\end{array}
\right. 
\end{equation}
For the neutrino gas, the relativistic limit of Eq.\ (\ref{Cfqsmall}) yields
\begin{equation}\label{Cnu}
C_\nu = 
-\frac{1}{24\pi^2}\int_0^\infty d{\cal P}
(f_\nu({\cal P}) - f_{\bar\nu}({\cal P})) \,,
\end{equation}
while $A_\nu$ and $B_\nu$ are given in Eq.\ (\ref{ABnuqsmall}).

With the help of these formulas and remembering that 
$\mbox{Re}\,\pi^{(m)}_T = 4e^2\omega^2_{0e}$ in the case we are considering,
the dispersion relation becomes
\begin{equation}\label{tdisprel2}
q^2 = 4e^2\omega^2_{0e} + 2T^2_0\omega^2_{0\nu}\left[
1 \mp \frac{{\cal Q} T^\prime_0}{2m_e T_0}\right]^2
\left[1 \mp \frac{C_\nu{\cal Q}}{\omega^2_{0\nu}}
\left(\frac{\tilde u\cdot u^\prime}{\tilde u^2}\right)\right]^2
\pm \frac{\Pi_P^{(\nu)}{\cal Q}}{{\cal Q}^\prime}
\left(\frac{\tilde u\cdot u^\prime}{\tilde u^2}\right) \,,
\end{equation}
where we have included the $\Pi_P^{(\nu)}$ term as indicated
in Eq.\ (\ref{pisubpinu}).

Let us consider first the situation in which
$f_\nu \approx f_{\bar\nu}$, so that $\Pi_P^{(\nu)}$ and
$C_\nu$ can be neglected in Eq.\ (\ref{tdisprel2}). The solutions for
the two modes are then given by
\begin{equation}\label{sol1}
\Omega_\pm = \sqrt{{\cal Q}^2 + 4e^2\omega^2_{0e}} \mp 
\frac{T_0 T^\prime_0\omega^2_{0\nu}}{m_e}
\frac{{\cal Q}}{\sqrt{{\cal Q}^2 + 4e^2\omega^2_{0e}}} \,,
\end{equation}
which is of the same form as that given in Eq.\ (4.22) of Ref.\ \cite{pisubpi2},
if we make the correspondence
\begin{equation}\label{equiv1}
\frac{1}{2}aR_\nu \rightarrow \frac{T_0 T^\prime_0\omega^2_{0\nu}}{m_e}
\end{equation}
there.  For the situations of potential interest analyzed in Ref.\ \cite{pisubpi2},
the effects of the dispersion relations given in Eq.\ (\ref{sol1}) are
smaller than those found in that reference by a factor of about
$G_F\omega^2_{0e} \approx G_F n_e/m_e$, and whence are unimportant
for all practical purposes.

Therefore, retaining only the term proportional to $\Pi^{(\nu)}_P$ in
Eq.\ (\ref{tdisprel2}), the dispersion relation becomes
\begin{equation}\label{tdisprelnu}
q^2 = 4e^2\omega^2_{0e} \pm \xi\Pi_P^{(\nu)} \,,
\end{equation}
where
\begin{equation}\label{xi}
\xi = \frac{{\cal Q}}{{\cal Q}^\prime}
\left(\frac{\tilde u\cdot u^\prime}{\tilde u^2}\right) \,.
\end{equation}
Of course, when the neutrino gas is not moving relative to the
matter background ($\vec U^\prime = 0$), $\xi = 1$ and Eq.\ (\ref{tdisprelnu})
reduces to the form given in Ref.\ \cite{pisubpi2}.

The salient feature here is that, in general, the dispersion
relation is not isotropic, so that the splitting between the two
circularly polarized modes is different depending on the direction of
propagation of the photon relative to the velocity of the neutrino gas.
To assess the consequences that this effect can have 
on the analysis given in Ref.\ \cite{pisubpi2} we consider two
extreme cases.

\paragraph{$\vec{\cal Q}$ perpendicular to $\vec U^\prime$.}

Using $\vec{\cal Q}\cdot\vec U^\prime = 0$, it is very simple
to verify that 
\begin{displaymath}
\tilde u\cdot u^\prime = -\frac{U^{\prime\,0}{\cal Q}^2}{q^2} \,,
\end{displaymath}
while ${\cal Q}^\prime = \sqrt{\vec U^{\prime\,2} + {\cal Q}^2}$. Using
$\tilde u^2 = -{\cal Q}^2/q^2$, it then follows that
\begin{equation}\label{xiperp}
\xi = \frac{{\cal Q}\sqrt{1 + \vec U^{\prime\,2}}}
{\sqrt{\Omega^2\vec U^{\prime\,2} + {\cal Q}^2}} \,.
\end{equation}
For small velocities of the neutrino gas this reduces to 1, as it should be,
while for large velocities it implies that the effect of the $\Pi^{(\nu)}_P$
term is reduced by the factor ${\cal Q}/\Omega$ for $\Omega > {\cal Q}$.

\paragraph{$\vec{\cal Q}$ parallel to $\vec U^\prime$.}

We set
\begin{equation}\label{Qparallel}
\vec Q = \lambda{\cal Q}\hat U^\prime
\end{equation}
to include the possibility that the photon propagates antiparallel to
the velocity of the neutrino gas. A little algebra shows that in this case
\begin{displaymath}
\tilde u\cdot u^\prime = \frac{-U^{\prime\,0}{\cal Q}^2 + 
\lambda\Omega{\cal Q}|\vec U^\prime|}{q^2}
\end{displaymath}
while ${\cal Q}^\prime = \left|\Omega|\vec U^\prime| - 
\lambda{\cal Q} U^{\prime\,0}\right|$. Therefore
\begin{equation}\label{xipar}
\xi = \frac{{\cal Q} - \lambda \beta^\prime\Omega}
{\left|\Omega\beta^\prime - \lambda{\cal Q} \right|} \,,
\end{equation}
where we have defined the speed of the neutrino gas
\begin{equation}\label{nugasvel}
\vec\beta^\prime = \frac{\vec U^\prime}{U^{\prime\,0}} \,.
\end{equation}

Eq.\ (\ref{xipar}) reveals in a simple way the anisotropic nature of the dispersion
relation. For example, if the velocity of the neutrino stream is such that
\begin{equation}\label{slowphoton}
\frac{{\cal Q}}{\Omega} < \beta^\prime \,,
\end{equation}
then $\xi = -1$ or $+1$ depending on whether the photon is propagating
parallel or antiparallel to $\vec\beta^\prime$, respectively.
In particular, this implies that the frequency difference between the 
dispersion relations of the two (circularly polarized) transverse modes
is the opposite to what it is if the neutrino gas is not moving relative
to the matter background. This effect is easy to understand by noticing
that, if the condition in Eq.\ (\ref{slowphoton}) holds, then a photon
moving parallel to $\vec\beta^\prime$  appears to be moving in the opposite
direction in the rest frame of the neutrino gas.

%
%
\section{Outlook}
\label{sec:conclusions}

Although our work has been largely motivated by the study of
the electromagnetic properties of a neutrino gas that moves, as
a whole, relative to a plasma, the approach is applicable to
a wider class of problem in similar physical systems,
involving relativistic plasmas or high temperature gauge theories.
The field theory methods employed here allow us to consider
situations for which the semiclassical approaches, such as those
based on kinetic theory, are not suitable, and those
for which the full power of the  
techniques and methods that have been developed 
for high temperature field theory calculations
must be employed.
Some possible extensions of the present work along those lines
would involve the calculation of the imaginary part
of the self-energy to determine the damping rates,
and the application of the resummation methods\cite{resumm} 
to study the dispersion relations 
in those circumstances in which the perturbative
approximations are not reliable.

\paragraph*{Acknowledgement}

This work has been partially supported by the
U.S. National Science Foundation Grant PHY-9900766.


\begin{thebibliography}{11}
\bibitem{footnote1} Our own results, discussed in 
Section \ref{subsec:nuinstabilities},
do not support the conclusions of Ref.\ \cite{silvaetal}.
\bibitem{footnote2} In Ref.\ \cite{pisubpi} $R^{(\pm)}_{\alpha\beta}$ were
denoted by $A_{\alpha\beta}$ and $B_{\alpha\beta}$, respectively.
\end{thebibliography}
\end{document}